\documentclass{llncs}
\usepackage{makeidx}  

\usepackage{graphicx}
\usepackage{wrapfig}
\usepackage{sidecap}
\usepackage{cite}

\begin{document}
%
%
%
%
%
\mainmatter              
\title{Hidden attractors in \\ fundamental problems and engineering models.
\thanks{
International Conference on Advanced Engineering - Theory and Applications,
2015 (Ho Chi Minh City, Vietnam), \url{http://icaeta.org/}, plenary lecture}}
\titlerunning{Hamiltonian Mechanics}  
%
\author{Nikolay V. Kuznetsov}
\authorrunning{Nikolay V. Kuznetsov} 
%
\tocauthor{Nikolay Kuznetsov}
\institute{
Faculty of Mathematics and Mechanics,
Saint-Petersburg University, \\ St. Petersburg 198504, Russia,
\and
Department of Mathematical Information Technology,
University of Jyv\"{a}skyl\"{a}, \\
40014 Jyv\"{a}skyl\"{a}, Finland\\
\email{nkuznetsov239@gmail.com}
}

\maketitle              

\begin{abstract}
 Recently a concept of self-excited and hidden attractors was suggested:
 an attractor is called a \emph{self-excited attractor} if its basin of attraction
 overlaps with neighborhood of an equilibrium, otherwise it is called a \emph{hidden attractor}.
 For example, hidden attractors are attractors in systems
 with no equilibria or with only one stable equilibrium
 (a special case of multistability and coexistence of attractors).
 While coexisting self-excited attractors
 can be found using the standard computational procedure,
 there is no standard way of predicting the existence
 or coexistence of hidden attractors in a system.

 In this  plenary survey lecture the concept of self-excited and hidden attractors is discussed,
 and various corresponding examples of self-excited and hidden attractors are considered.
 The material is mostly based on 
 surveys \cite{LeonovK-2011-IFAC,LeonovK-2013-IJBC,KuznetsovL-2014-IFACWC,LeonovKM-2015-EPJST}.
\end{abstract}
\section{Analytical-numerical study of oscillations}

 An oscillation dynamical system can generally be easily numerically localized if the initial data from its open
 neighborhood in the phase space (with the exception of a minor set of points) lead to
 a long-term behavior that approaches the oscillation.
 Such an oscillation (or a set of oscillations)
 is called an attractor, and its attracting set is called the basin of attraction
 (i.e., a set of initial data for which the trajectories tend to the attractor).

 When the theories of dynamical systems, oscillations, and chaos were first developed
 researchers mainly focused on analyzing equilibria stability,
 which can be easily done numerically or analytically,
 and on the birth of periodic or chaotic attractors
 from unstable equilibria.
 The structures of many physical dynamical systems
 are such that it is almost obvious that attractors exist
 because the trajectories can not tend to infinity
 and the oscillations are excited by an unstable equilibrium
 (see, e.g., the Rayleigh \cite{Rayleigh-1877}, Duffing \cite{Duffing-1918}, van der Pol \cite{VanDerPol-1926},
 Tricomi \cite{Tricomi-1933}, Beluosov-Zhabotinsky \cite{Belousov-1959}, and Lorenz \cite{Lorenz-1963} systems).
 This meant that scientists of that time could compute such attractors
 by constructing a solution using initial data from a
 small neighborhood of the equilibrium, observing how it is attracted and, thus, visualizes the attractor.
 In this \emph{standard computational procedure}, computational methods and the engineering notion of a
 \emph{transient process} were combined to study oscillations.

\section{Self-excited and hidden attractors}

 From a computational perspective, it is natural to suggest the following classification
 of attractors,
 which is based on the connection of their basins of attraction with equilibria in the phase space:

\begin{definition}
\cite{LeonovKV-2011-PLA,LeonovKV-2012-PhysD,LeonovK-2013-IJBC,LeonovKM-2015-EPJST}
 An attractor is called a \emph{self-excited attractor}
 if its basin of attraction
 intersects with any open neighborhood of a stationary state (an equilibrium),
 otherwise it is called a \emph{hidden attractor}.
\end{definition}

\begin{figure}[!ht]
\centerline{%
\begin{tabular}{c c c}
\includegraphics[width=0.33\textwidth]{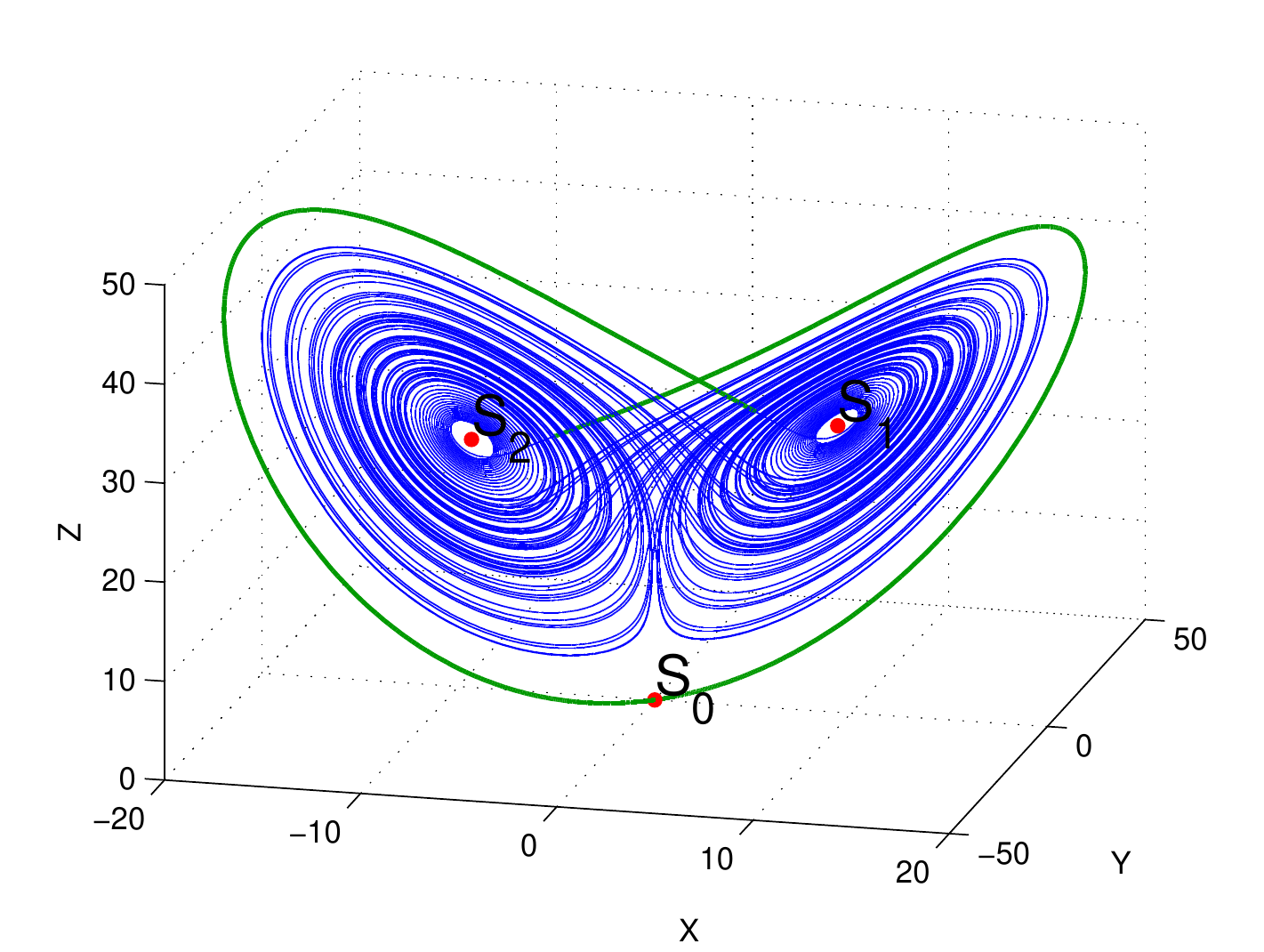} &
\includegraphics[width=0.33\textwidth]{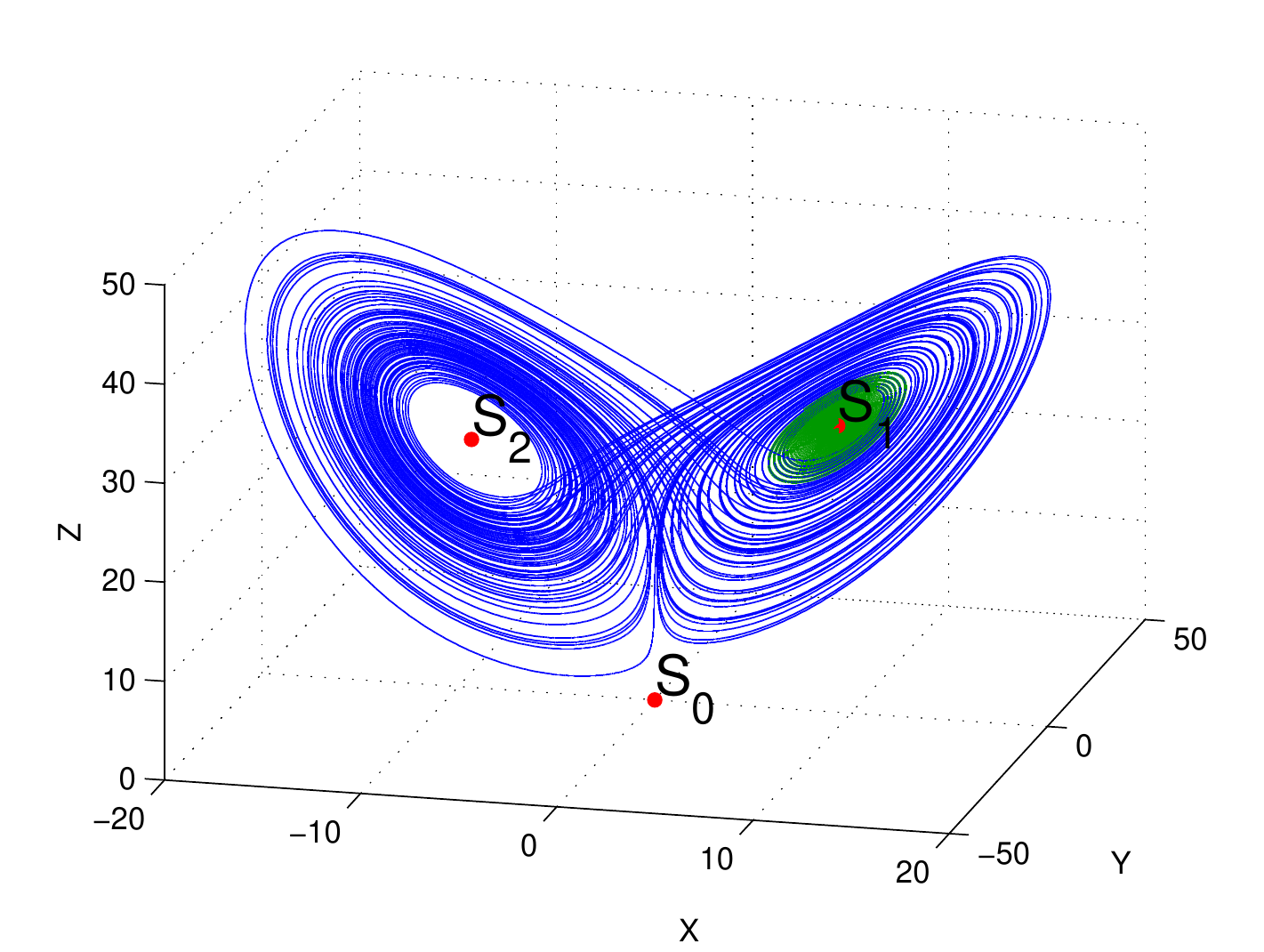} &
\includegraphics[width=0.33\textwidth]{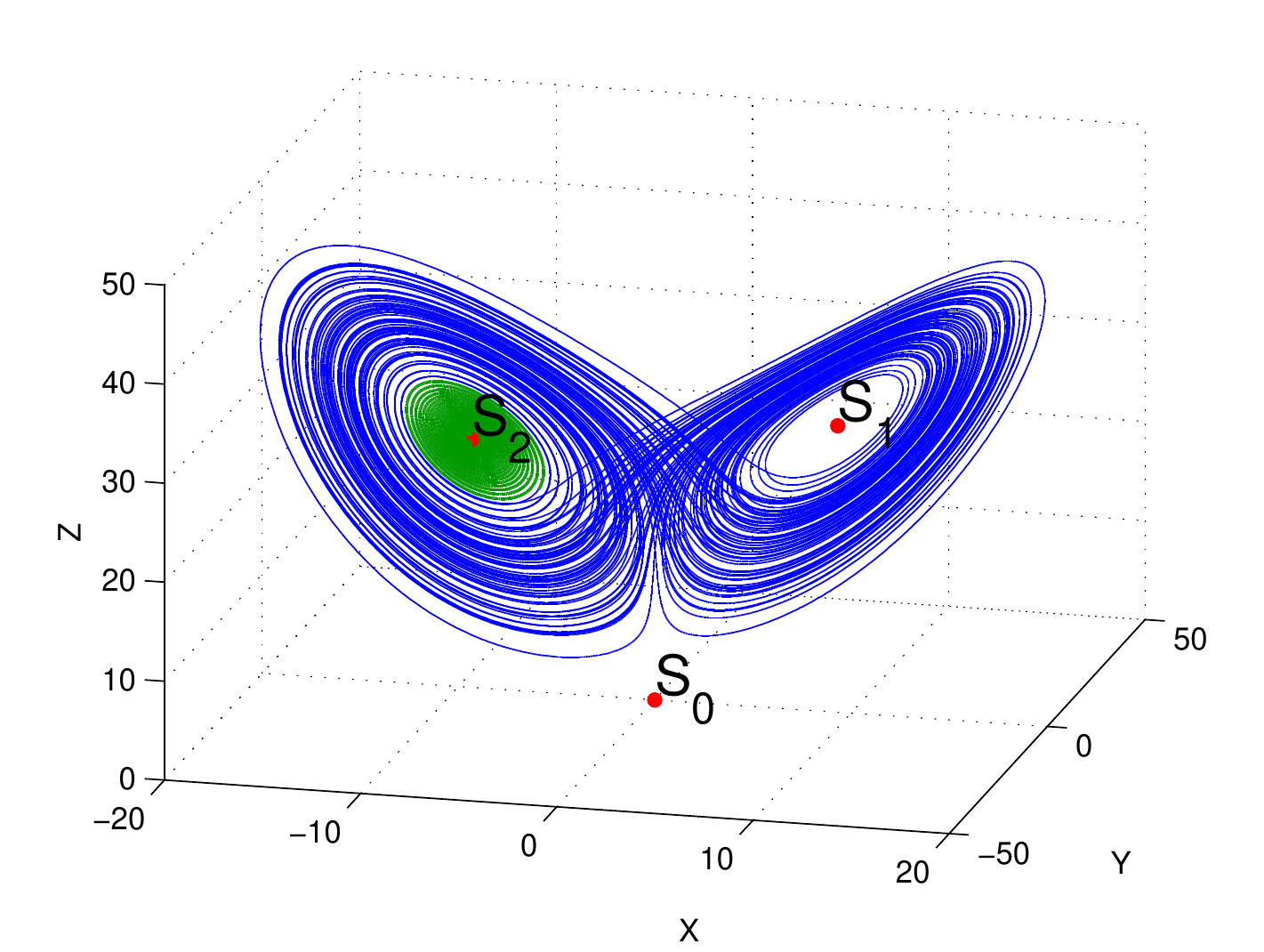}
\end{tabular}}
 \caption{
 Numerical visualization of
 the classical self-excited chaotic attractor in the Lorenz system
 $
 \dot x= 10(y-x),
 \dot y= 28 x-y-xz,
 \dot z=-8/3 z+xy.
 $
 The attractor is self-excited respect to all three equilibria:
 it can be visualized by trajectories that start in small neighborhoods
 of any of unstable equilibria $S_{0,1,2}$.
 Here the separation of trajectory into transition process (green)
 and approximation of attractor (blue) is rough.
 }
\label{attr-lorenz-3unstable}
\end{figure}

 The first well-known example of a visualization of chaotic attractor
 in a dynamical system from the work of Lorenz \cite{Lorenz-1963}
 corresponds to the excitation of chaotic attractor
 from unstable equilibria.
 For classical parameters the Lorenz attractor
 is self-excited with respect to all three equilibria
 and could have been found using the standard computational procedure
 (see Fig.~\ref{attr-lorenz-3unstable}).
 Note that the chaotic attractor in the Lorenz system with other parameters 
 may be self-excited with respect to zero unstable equilibrium only,
 and the possible existence in the Lorenz system of a hidden chaotic attractor is an open problem.

\begin{figure}[!ht]
\centerline{%
\begin{tabular}{c c}
\includegraphics[width=0.48\textwidth]{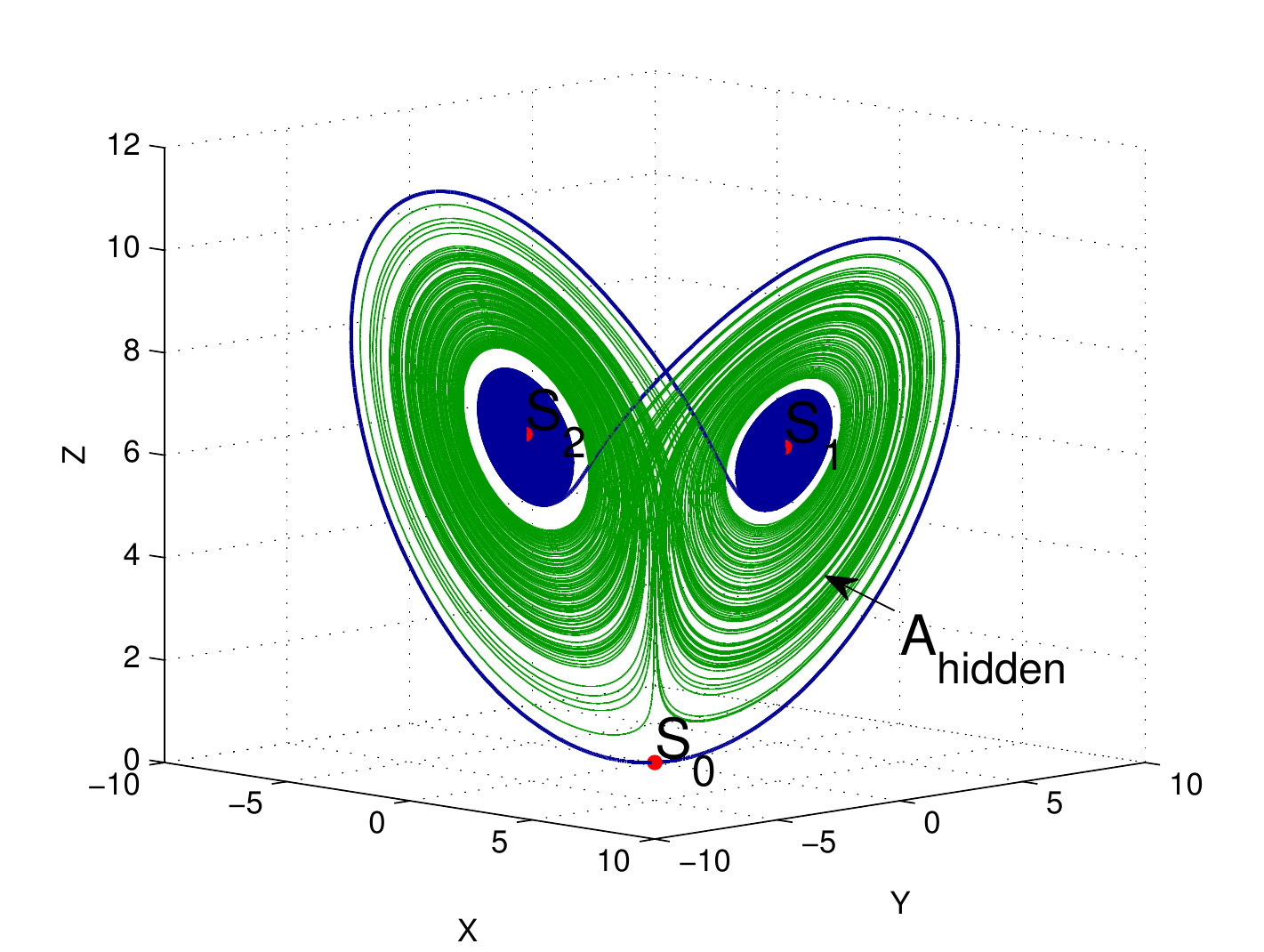} &
\includegraphics[width=0.48\textwidth]{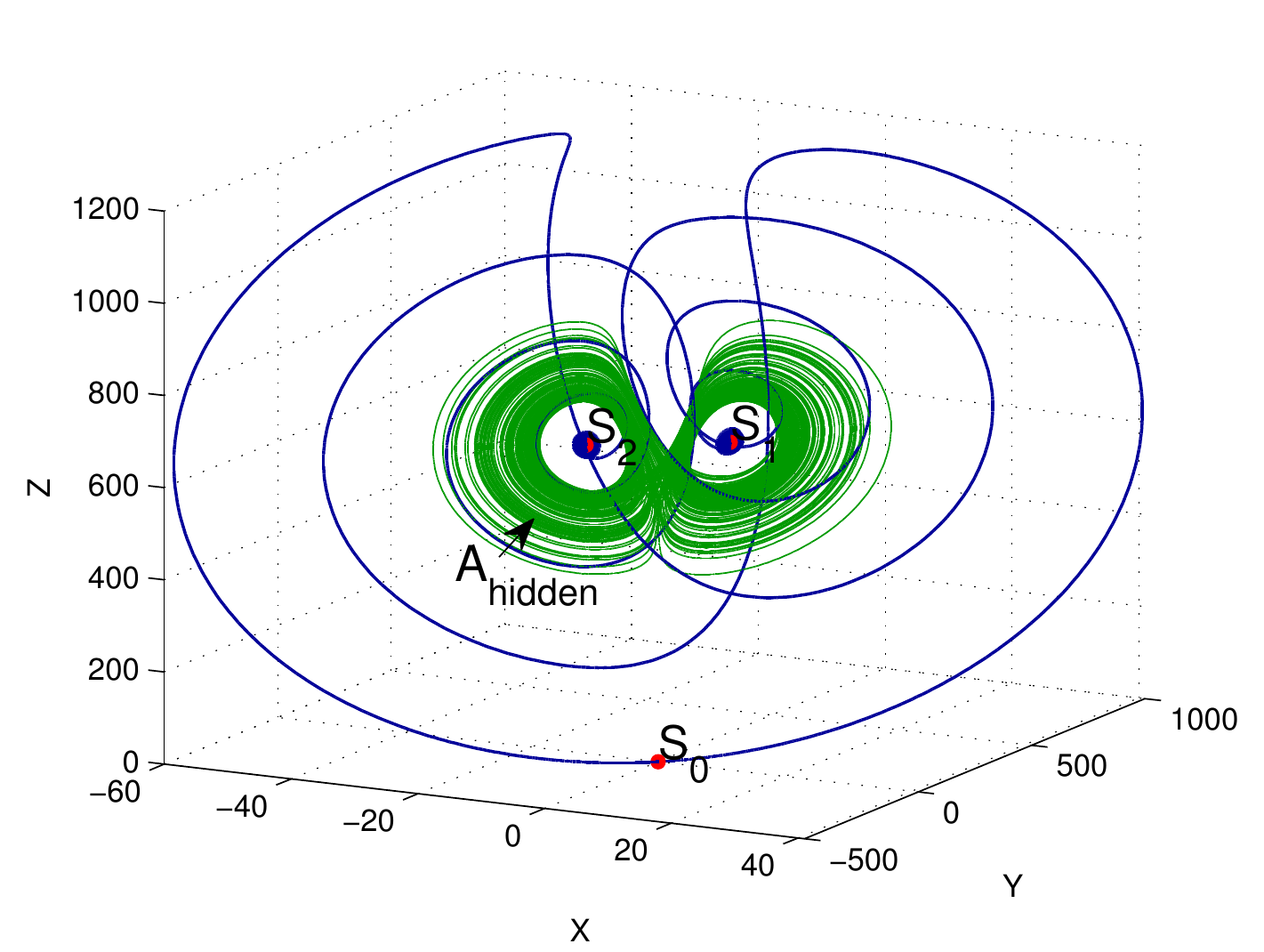} \\
 a.~~
 $r = 6.8, a = -0.5, \sigma=ra, b=1$.
 &
 b.~~
 $r = 700, a = 0.0052, \sigma=ra, b=1$.
 \end{tabular}}
 \caption{
 Numerical visualization of hidden attractor (green trajectory) in the generalized Lorenz system.
 Outgoing separatrices of the zero saddle equilibrium
 are attracted to the stable equilibria  $S_{1,2}$ (blue trajectories).
 }
\label{attr-rab-gl-hidden}
\end{figure}

 At the same time in the generalized Lorenz system
 $
 \dot x = -\sigma(x - y) - ayz,
 \dot y = r x - y - xz,
 \dot z = -bz + xy
 $
 hidden chaotic attractors can be found \cite{LeonovKM-2015-CNSNS,KuznetsovLM-2015,LeonovKM-2015-EPJST}
 (see Fig.~\ref{attr-rab-gl-hidden}).
 For negative $a<0$ the system corresponds to the Rabinovich system, which describes the interaction of plasma waves
 and was considered in 1978 \cite{Rabinovich-1978,PikovskiRT-1978};
 for positive $a>0$ it corresponds to the Glukhovsky-Dolghansky system,
 which describes convective fluid motion and was considered in 1980 \cite{GlukhovskyD-1980};
 also it describes a rigid body rotation in a resisting medium and
 the forced motion of a gyrostat (see \cite{LeonovB-1992}).

The basin of attraction for a hidden attractor is not connected with any equilibrium.
For example, hidden attractors are attractors in systems
with no equilibria
or with only one stable equilibrium (a special case of the multistability and coexistence of attractors).
Note that multistability
can be undesired in various practical applications.
At the same time the coexisting self-excited attractors in multistable systems
(see, e.g. various examples of multistable engineering systems in famous book \cite{AndronovVKh-1937},
and recent physical examples in \cite{PisarchikF-2014})
can be found using the standard computational procedure,
whereas there is no standard way of predicting the existence
or coexistence of hidden attractors in a system.

\begin{figure}[!ht]
 \centering
 \includegraphics[width=0.45\textwidth]{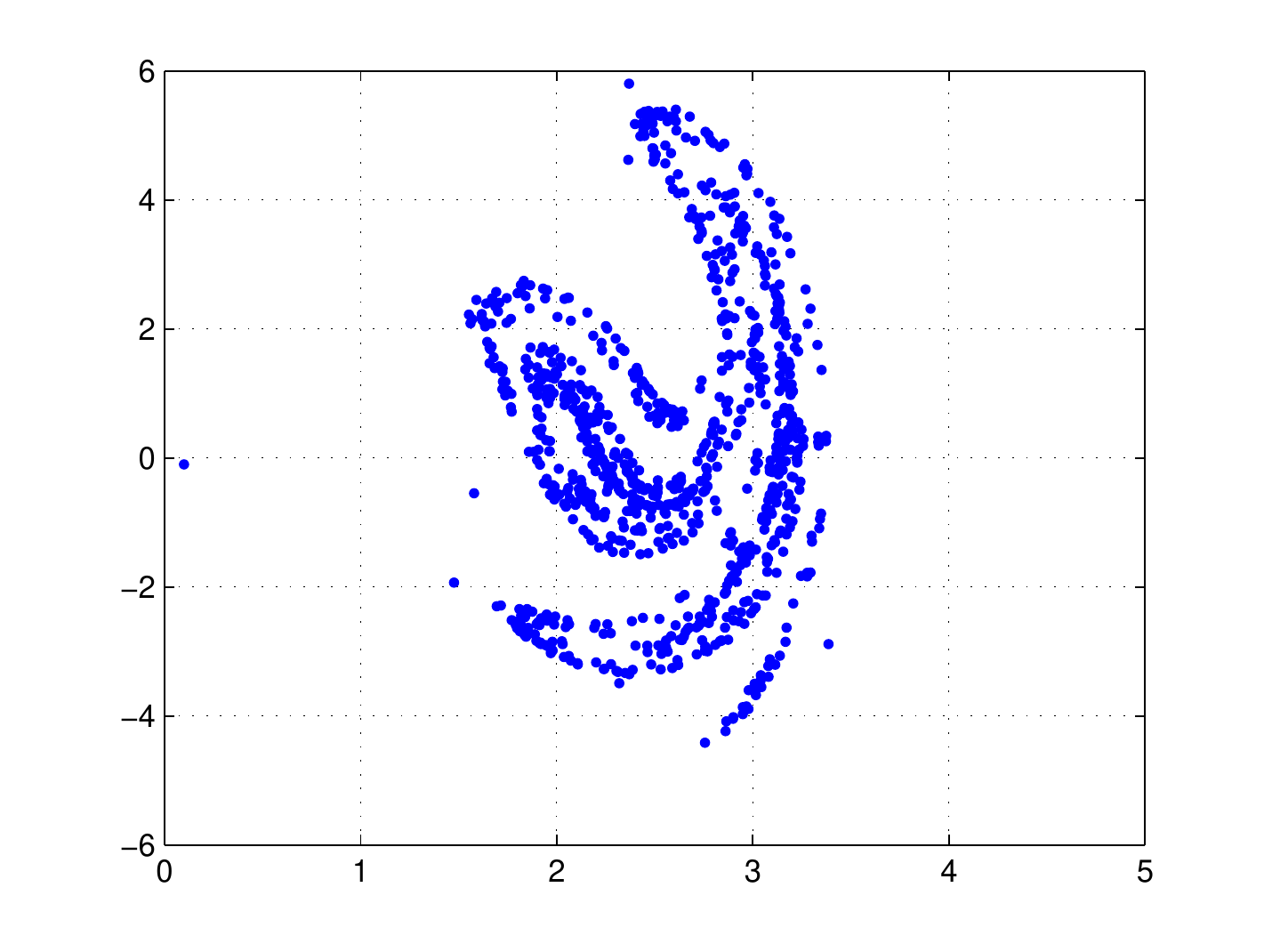}
 \caption{
   An attractor in forced Duffing oscillator $\ddot x+0.05\dot x+x^3=7.5\cos(t)$
   is self-excited with respect to the zero equilibrium of
   unperturbed system (without $7.5\cos(t)$).
 }
 \label{fig:uead}
\end{figure}

For nonautonomous systems, depending on the physical problem statement,
the notion of self-excited and hidden attractors can be introduced
with respect to the stationary states
of the system at the fixed initial time or the corresponding system without time-varying excitations.
For example, one of the classical examples of self-excited chaotic attractors
was numerically found by Ueda in 1961 \cite{UedaAH-1973}
in a forced Duffing system $\ddot x+0.05\dot x+x^3=7.5\cos(t)$.
To construct a self-excited chaotic attractor in this system (Fig.~\ref{fig:uead})
it was used a transient process from the zero equilibrium
of the unperturbed autonomous system (i.e., without $\cos(t)$)
to an attractor in the forced system.
If the discrete dynamics of system are considered on a Poincare section,
then we can also use stationary or periodic points on the section that corresponds
to a periodic orbit of the system (the consideration of periodic orbits is also natural
for discrete systems).
Note that if the attracting domain is the
whole state space, then the attractor can be visualized by
any trajectory and the only difference between computations
is the time of the transient process.

One of the first well-known problems of analyzing hidden periodic oscillations
arose in connection with the second part of Hilbert's 16th problem (1900) \cite{Hilbert-1901}
on the number and mutual disposition of limit cycles
in two-dimensional polynomial systems
(see, e.g. resent results \cite{LeonovK-RCD-2010,LeonovKKSV-2011-WTSC,KuznetsovKL-2013-DEDS,LeonovK-2013-IJBC}
on visualization of nested limit cycles in quadratic systems:
$\dot x = \alpha_1x+\beta_1y+a_1x^2+b_1xy+c_1y^2,\ \dot y = \alpha_2x+\beta_2y+a_2x^2+b_2xy+c_2y^2$).

Later, in the 1950s-1960s, the study of the well-known
Aizerman's and Kalman's conjectures on absolute stability
led to the discovery of the possible coexistence of
a hidden periodic oscillation and a unique stable stationary point
in automatic control systems.
In 1957 R.E.~Kalman stated the following \cite{Kalman-1957}:
\begin{wrapfigure}{r}{0.5\textwidth}
 \centering
 \includegraphics[width=0.45\textwidth]{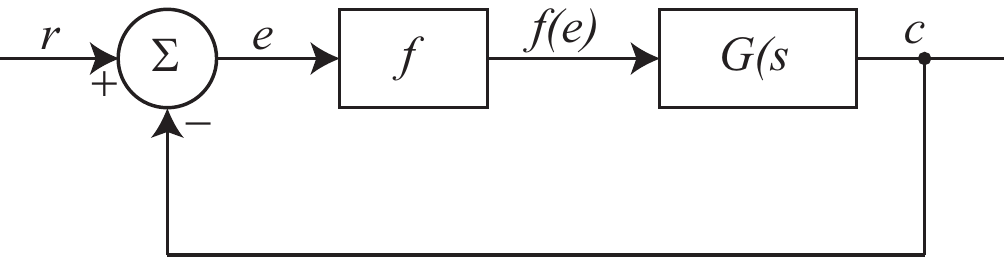}
 \caption{Nonlinear control system.
 $G(s)$ is a linear transfer function, $f(e)$ is a single-valued,
continuous, and differentiable \cite{Kalman-1957}}
 \label{KalmanScheme}
\end{wrapfigure}
``{\it
    If $f(e)$ in Fig.~1 {\rm [see Fig.~\ref{KalmanScheme}]} is
    replaced by constants $K$ corresponding
    to all possible values of $f'(e)$, and it is found that
    the closed-loop system is stable for all such $K$,
    then it is intuitively clear that the system must be monostable;
    i.e., all transient solutions will converge to a unique,
    stable critical point.}''
 Kalman's conjecture is a strengthening of Aizerman's conjecture \cite{Aizerman-1949},
 which considers nonlinearities belonging to the sector of linear stability.
 Note that these conjectures are valid from the standpoint of
 simplified analysis such as the linearization, harmonic balance, and describing function methods (DFM),
 what explains why these conjectures were put forward.
 Nowadays various counterexamples to these conjectures
 (nonlinear systems where the only equilibrium, which is stable,
 coexists with a hidden periodic oscillation) are known
 (see \cite{Pliss-1958,Fitts-1966,Barabanov-1988,BernatL-1996,LeonovBK-2010-DAN,BraginVKL-2011,LeonovK-2011-DAN,KuznetsovLS-2011-IFAC};
 the corresponding discrete examples are considered in \cite{Alli-Oke-2012-cu,HeathCS-2015}).

 Similar situation  with linear stability and hidden oscillations
 occur in the analysis of aircrafts and launchers control systems with saturation \cite{AndrievskyKLP-2013-IFAC,AndrievskyKLS-2013-IFAC}.
In \cite{LauvdalMF-1997} the crash of aircraft YF-22 Boeing in April 1992\footnote{
\url{http://www.youtube.com/watch?v=M6sy-fxIhF0}},
caused by the difficulties of rigorous analysis and design of nonlinear control systems with saturation,
is discussed and the conclusion is made
that {\it``since stability in simulations does not imply stability of the physical control
system (an example is the crash of the YF22), stronger theoretical understanding is required''}.

\begin{figure}[h]
\centering
 \includegraphics[width=0.95\textwidth]{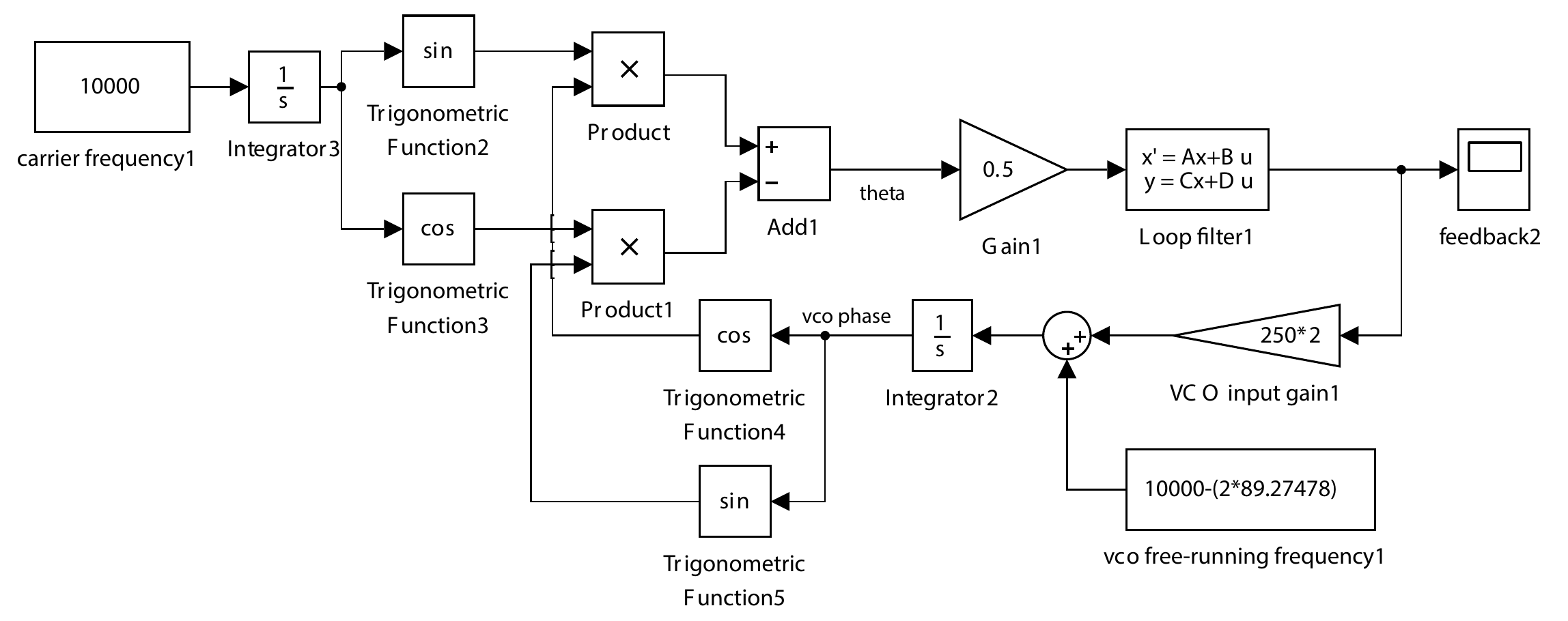}
 \caption{Model of two-phase PLL with lead-lag filter in MATLAB Simulink.
  Lead-lag loop filter with the transfer function
  $H(s) = \frac{1+s \tau_2}{1+s(\tau_1 + \tau_2)}$,
  $\tau_1 = 0.0448$, $\tau_2 = 0.0185$
  and the corresponding parameters
  $A = -\frac{1}{\tau_1+\tau_2}$,
  $B = 1 - \frac{\tau_2}{\tau_1+\tau_2}$,
  $C = \frac{1}{\tau_1+\tau_2}$,
  $D = \frac{\tau_2}{\tau_1+\tau_2}$.
 }
 \label{simulink-2phase-pll-model}
\end{figure}

\begin{figure}[!ht]
  \centering
  \includegraphics[width=0.5\linewidth]{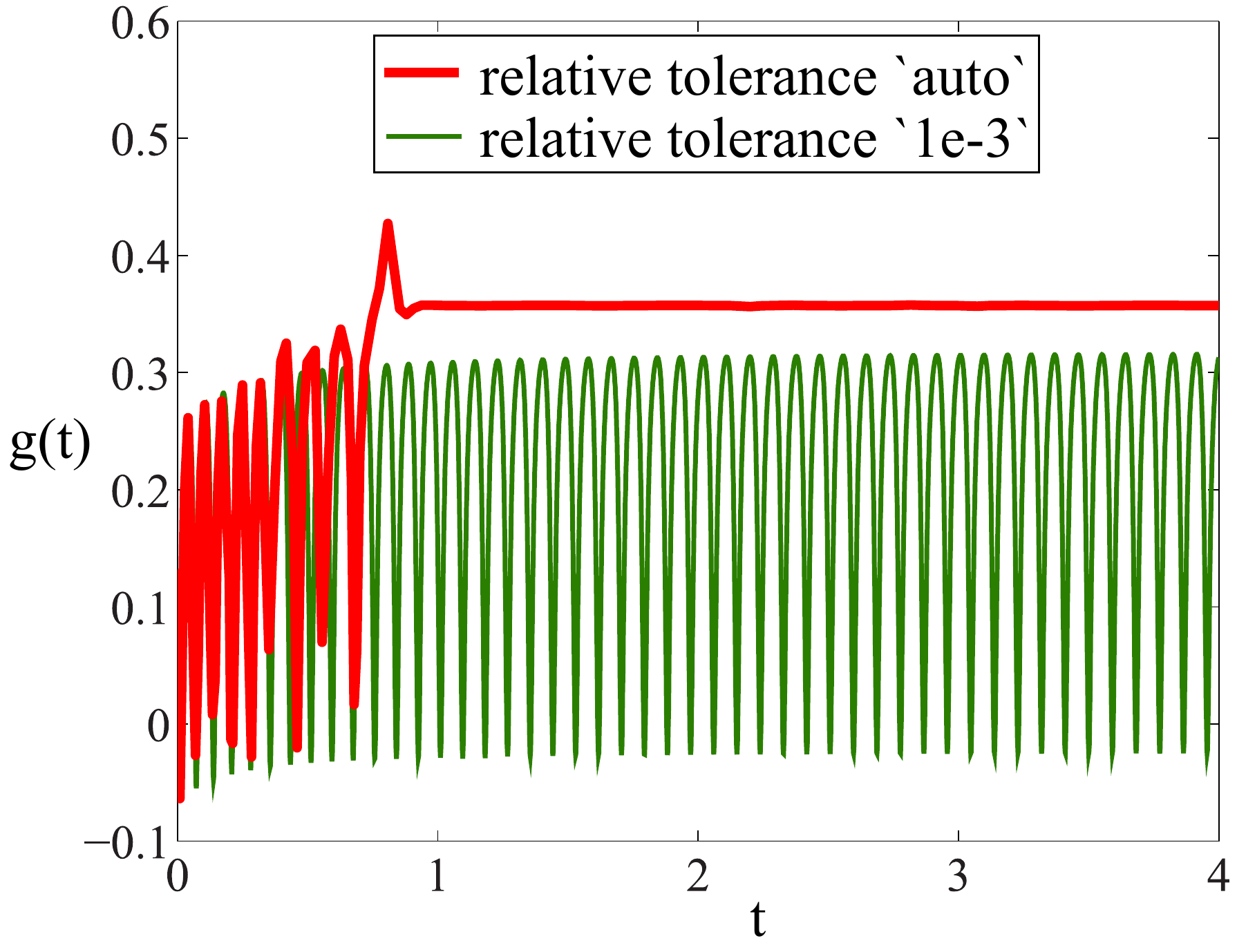}
  \caption{
  Simulation of two-phase PLL.
  Loop filter output $g(t)$ for the initial data $x_0\!=\!0.1318$
  obtained for default ``auto'' relative tolerance (red) --- acquires lock,
  relative tolerance set to ``1e-3''(green) --- does not acquire lock.}
  \label{pll_hidden}
\end{figure}

Corresponding limitations, caused by hidden oscillations, appear
in simulation of various phase-locked loop (PLL) based systems
\cite{LeonovK-2013-IJBC,KuznetsovLYY-2014-IFAC,KuznetsovKLNYY-2014-ICUMT-QPSK,KuznetsovKLSYY-2014-ICUMT-BPSK,KudryashovaKKLSYY-2014-ICINCO,KuznetsovKLNYY-2015-ISCAS,BestKKLYY-2015-ACC,BianchiKLYY-2015,LeonovKYY-2015-TCAS}.
PLL was designed to synchronize the phases of local oscillator and reference oscillator signals.
Next example shows that the use of default simulation parameters in MATLAB Simulink
for the study of two-phase PLL in the presence of hidden oscillation (see Fig.~\ref{simulink-2phase-pll-model})
can lead to the conclusions concerning the stability of the loop
and the pull-in (or capture) range\footnote{See discussion of rigorous definitions in \cite{KuznetsovLYY-2015-IFAC-Ranges,LeonovKYY-2015-TCAS}.}.
In Fig.~\ref{pll_hidden}
the model in Fig.~\ref{simulink-2phase-pll-model} simulated with relative tolerance set to ``1e-3''
or smaller does not acquire lock (black color),
but the model with default parameters (a relative tolerance set to ``auto'')
acquires lock (red color).
The same problems are also observed in SIMetrics SPICE model \cite{KuznetsovLYY-2014-IFAC,KuznetsovKLNYY-2015-ISCAS,BestKKLYY-2015-ACC}.
From a mathematical point of view, the above case corresponds to the existence of semistable periodic trajectory
or co-existence of unstable and stable periodic trajectories (here the stable one is a hidden oscillation)
\cite{Shakhtarin-1969,Belyustina-1970-eng,LeonovK-2013-IJBC,KuznetsovLYY-2014-IFAC}.
Therefore, if the gap between stable and unstable periodic trajectories
is smaller than the discretization step,
the numerical procedure may slip through the stable trajectory \cite{KuznetsovKLNYY-2015-ISCAS,BianchiKLYY-2015}.

Attractor in the \emph{systems without equilibria} are the hidden attractors, according to the above definition.
Systems without equilibria and with hidden oscillations appear naturally in the study of
various electromechanical models with rotation and electrical circuits with cylindrical phase space.
One of the first examples is from paper \cite{Sommerfeld-1902}, published in 1902,
in which Sommerfeld analyzed the vibrations caused by a motor driving an unbalanced weight
and discovered the so-called Sommerfeld effect (see, e.g., \cite{BlekhmanIF-2007,Eckert-2013}).
Another well-known chaotic system with no equilibrium points
is the Nos\`{e}--Hoover oscillator \cite{Nose-1984,Hoover-1985,Sprott-1994}.
An example of hidden chaotic attractor in electromechanical model with no equilibria
was reported in a power system in 2001 \cite{Venkatasubramanian-2001}.
Recent examples of hidden attractors in the systems without equilibria
can be found, e.g. in \cite{WeiWL-2014-HA, PhamJVWG-2014-HA, PhamRFF-2014-HA, PhamVG-2014-woeq, PhamVG-2014-woeq, PhamVJWW-2014,TahirJPVW-2015-woeq,VaidyanathanVP-2015-woeq, CafagnaG-2015-woeq,Chen-2015-IFAC-HA}.

\begin{SCfigure}
 \caption{
 Hidden chaotic attractor (green) in Chua circuit:
 $
  \dot x=\alpha(y-x-m_1x-\psi(x)),
  \dot y=x-y+z, \dot z=-(\beta y+\gamma z),
  \psi(x)=(m_0-m_1){\rm sat}(x).
 $
 Locally the stable zero equilibrium $F_0$ attracts trajectories (black)
 from stable manifolds $M^{st}_{1,2}$
 of two saddle points $S_{1,2}$;
 trajectories (red) from the unstable manifolds $M^{unst}_{1,2}$ tend to infinity;
 $\alpha = 8.4562,\ \beta = 12.0732,\ \gamma = 0.0052,\ m_0 = -
0.1768,\ m_1 = -1.1468$.}
\label{attr-Chua-hidden}
  \includegraphics[width=0.6\textwidth]{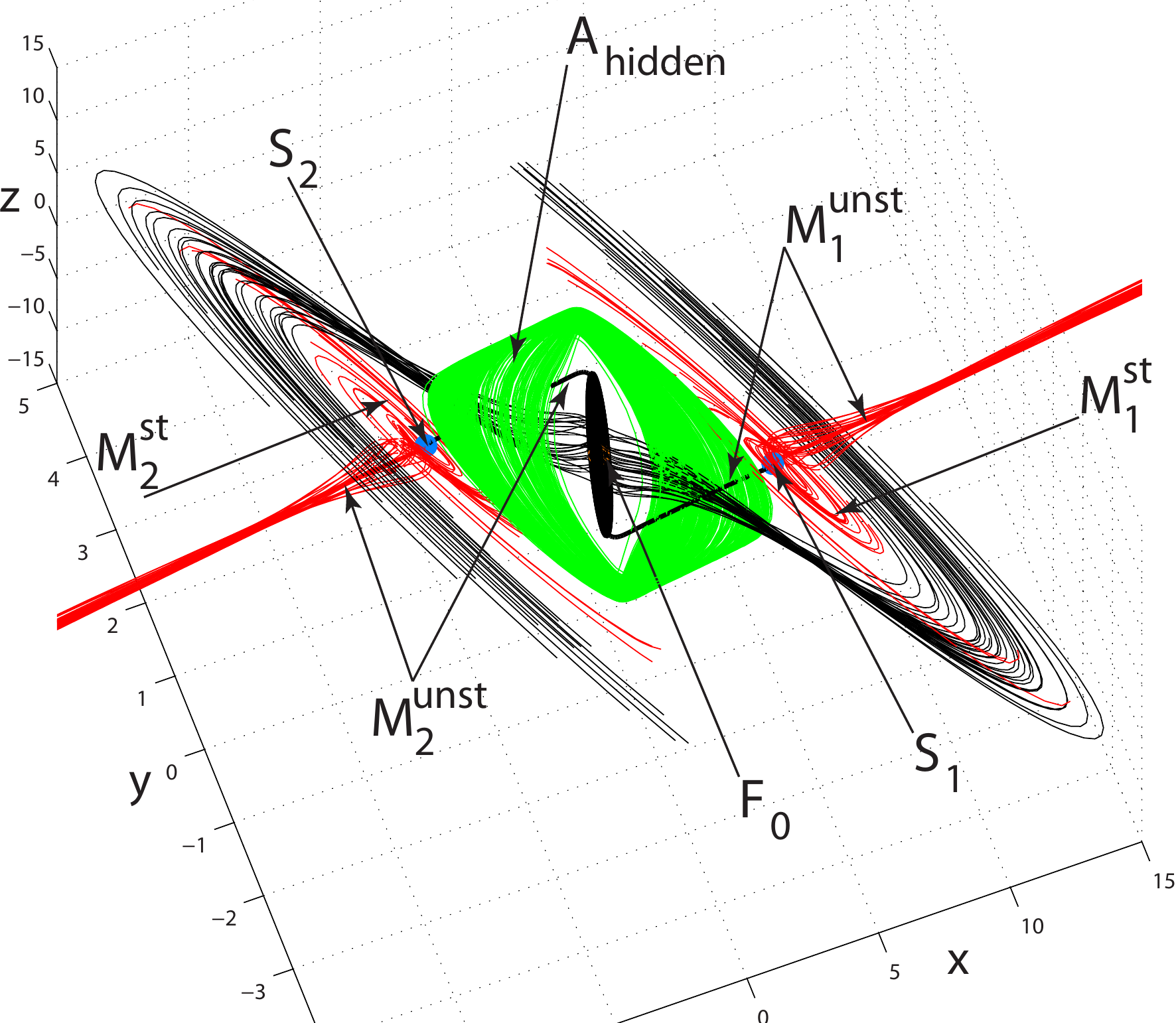}
\end{SCfigure}

After the concepts of hidden chaotic attractors was introduced first in connection with
discovery of the first hidden Chua attractors (see Fig.~\ref{attr-Chua-hidden})
\cite{LeonovK-2009-PhysCon,KuznetsovLV-2010-IFAC,KuznetsovVLS-2011,LeonovKV-2011-PLA,KuznetsovKLV-2011-ICINCO,LeonovKV-2012-PhysD,LeonovK-2012-IEEE,KuznetsovKLV-2013,LeonovKKK-2015-IFAC},
the hidden chaotic attractors have received much attention.
Recent examples of hidden attractors
can be found in
\cite{KiselevaKLN-2013,KiselevaKKLS-2014,LeonovKKSZ-2014,LeonovK-2014-JMS,LeonovK-2013-AISC,LeonovKKN-2013-JAND,KislevaKKL-2015-IFAC,
SprottWC-2013,WangC-2013,ZhusubaliyevM-2015-HA,WangSWZ-2015-HA,SharmaSPKL-2015,DangLBW-2015-HA,KuznetsovKMS-2015-HA,PhamVJWV-2014-HA,PhamJVWG-2014-HA,WeiWL-2014-HA,LiSprott-2014-HA,PhamRFF-2014-HA,WeiML-2014-HA,
PhamVVLV-2015-HA,ChenYB-2015-HA,ChenLYBXW-2015-HA,WeiZWY-2015-HA,BurkinK-2014-HA,WeiZ-2014-HA,LiZY-2014-HA,ZhaoLD-2014-HA,LaoSJS-2014-HA,ChaudhuriP-2014-HA,
PhamVJW-2014,PhamVJWW-2014,KingniJSW-2014,LiSprott-2014-PLA-cu,MolaieJSG-2013,JafariSPGJ-2014,JafariS-2013-cu,KingniJSW-2014,DudkowskiPK-2015-HA,WeiYZY-2015-HA,ZhusubaliyevMRN-2015-HA,BaoHCXY-2015-HA,DancaFKC-2015-IJBC}).

See also
\emph{The European Physical Journal Special Topics: Multistability: Uncovering Hidden Attractors}, 2015
(see \cite{ShahzadPAJH-2015-HA,BrezetskyiDK-2015-HA,JafariSN-2015-HA,ZhusubaliyevMCM-2015-HA,SahaSRC-2015-HA,Semenov20151553,FengW-2015-HA,Li20151493,FengPW-2015-HA,Sprott20151409,Pham20151507,VaidyanathanPV-2015-HA}).

\section{Acknowledgments}
  This  work  was supported by Russian Scientific Foundation (project 14-21-00041, sec.2)
  and Saint-Petersburg State University (6.38.505.2014, sec.1).


\newcommand{\noopsort}[1]{} \newcommand{\printfirst}[2]{#1}
  \newcommand{\singleletter}[1]{#1} \newcommand{\switchargs}[2]{#2#1}

\end{document}